# The influence of impurities on the charge carrier mobility of small molecule organic semiconductors

*Pascal Friederich* [1,2,3,*], *Artem Fediai* [1], *Jing Li* [4], *Anirban Mondal* [5], *Naresh B. Kotadiya* [5], *Franz Symalla* [6], *Gert-Jan A. H. Wetzelaer* [5], *Denis Andrienko* [5], *Xavier Blase* [4], *David Beljonne* [7], *Paul W. M. Blom* [5], *Jean-Luc Brédas* [3], *Wolfgang Wenzel* [1,*]

[1] *Institute of Nanotechnology, Karlsruhe Institute of Technology, Hermann-von-Helmholtz-Platz 1, 76344 Eggenstein-Leopoldshafen, Germany*
[2] *Department of Chemistry, University of Toronto, 80 St George St, Toronto, ON M5S 3H6, Canada*
[3] *Chemistry and Biochemistry, Georgia Institute of Technology, 901 Atlantic Dr NW, Atlanta, GA 30332-0400, USA*
[4] *University Grenoble Alpes and CNRS, Institut Néel, 25 Avenue des Martyrs, 38042 Grenoble, France*
[5] *Max Planck Institute for Polymer Research, Ackermannweg 10, 55128 Mainz, Germany*
[6] *Nanomatch GmbH, Hermann-von-Helmholtz-Platz 1, 76344 Eggenstein-Leopoldshafen, Germany*
[7] *Laboratory for Chemistry of Novel Materials, University of Mons, Place du Parc 20, 7000 Mons, Belgium*

*Corresponding authors: pascal.friederich@kit.edu, wolfgang.wenzel@kit.edu

**Abstract**

Amorphous organic semiconductors based on small molecules and polymers are used in many applications, most prominently organic light emitting diodes (OLEDs) and organic solar cells. Impurities and charge traps are omnipresent in most currently available organic semiconductors and limit charge transport and thus device efficiency. The microscopic cause as well as the chemical nature of these traps are presently not well understood. Using a multiscale model we characterize the influence of impurities on the density of states and charge transport in small-molecule amorphous organic semiconductors. We use the model to quantitatively describe the influence of water molecules and water-oxygen complexes on the electron and hole mobilities. These species are seen to impact the shape of the density of states and to act as explicit charge traps within the energy gap. Our results show that trap states introduced by molecular oxygen can be deep enough to limit the electron mobility in widely used materials.

**TOC**

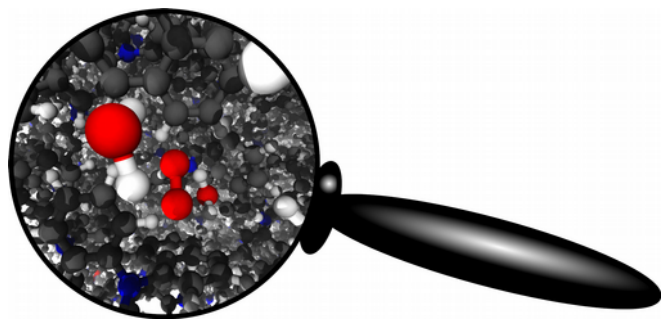

## 1. Introduction

Future prospects of organic light emitting diodes (OLEDs)[1,2] and organic solar cells[3,4] rely on the development of new organic semiconductors with optical and electronic properties outperforming those of presently available materials. High charge-carrier mobility and conductivity are required to achieve high device efficiencies. Trapping of charges by impurities can severely hinder charge transport and limit efficiencies. Thus, very pure and therefore costly materials are preferable.[5,6] The origin of charge trapping in organic-semiconductor materials is presently not well understood, in part because the chemical nature of the impurities is hard to characterize.[6–10] Computational materials design is becoming a widely used tool to complement and accelerate experimental efforts.[11–15] The possibility to virtually investigate microscopic processes in quantitative computational models was also shown to contribute to the understanding of experimentally observed phenomena.

For solution-processed conjugated polymers, it has been shown that electron transport is strongly hindered by the presence of electron traps.[5] Apart from slowing down charge mobility, these electron traps facilitate nonradiative trap-assisted recombination processes. The reduced electron transport and increased nonradiative recombination due to electron traps greatly compromise device performance in organic light-emitting diodes and organic solar cells.[16] It is therefore of fundamental importance to identify the electron trapping mechanism, which can possibly lead to strategies to remove the traps.[6,17]

In conjugated polymers, it was found that the electron traps are situated at a common energy below the vacuum level for a range of different materials, suggesting that the traps share a common origin.[5] The fact they are present in a range of polymers with different conjugated backbones suggests that the trap is caused by an impurity that is introduced during fabrication or commonly present in the environment, with the best candidates being oxygen or water. Based on the estimated energetic position, a $(H_2O)_2$-$O_2$ water complex was considered to be a likely candidate for the electron trap.[5,18,19] For vacuum-deposited organic small molecules, the effect of traps on electron transport has been studied to a lesser extent. In thermally evaporated films of bis(8-hydroxy-2-methylquinoline)-(4-phenylphenoxy)aluminum (BAlq)[20] and N,N'-di(1-naphthyl)-N,N'-diphenyl-(1,1'-biphenyl)-4,4'-diamine (α-NPD), electron transport could be described by incorporating electron-trapping sites with a concentration of $10^{24}$ m$^{-3}$, which is substantially higher than the typical charge concentrations under operating conditions in diodes, indicating that electron transport is heavily trap limited in these systems.[9,17]

Here, we investigate two potential trapping effects, broadening of the density of states (**Figure 1a**) and active trapping of charges (**Figure 1b**), and their influence on electron transport in three small-molecule materials. Using a multiscale simulation approach, we demonstrate that water molecules can broaden and introduce an exponential tail in the density of states, which significantly reduces the charge mobility in amorphous organic semiconductors. We further show that the previously suggested water-oxygen complex yields trap energy levels in these materials at deeper energy than originally anticipated.[17] We have quantitatively characterized the energylevels of a variety of other ubiquitous molecules that can act as traps and conclude that molecular oxygen is the most likely candidate for active trapping of electrons among the materials investigated.

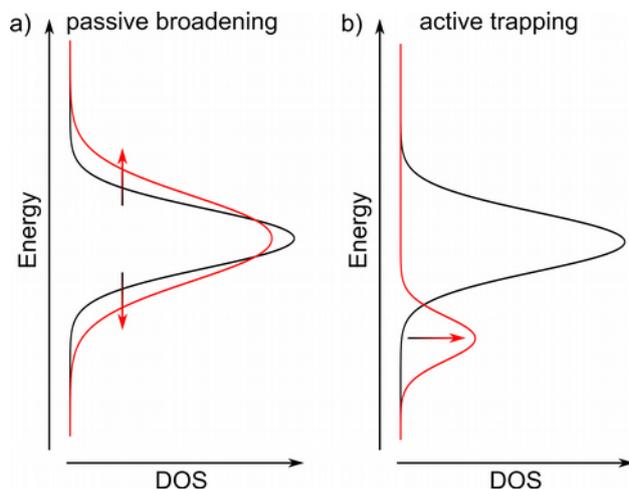

*Figure 1. Potential influence of impurities studied in this work. a) Passive broadening of the density of states (DOS) and exponential tail states caused by e.g. molecules with permanent dipole moments. b) Active trapping of charges caused by impurities with higher electron affinities than the host material.*

## 2. Methods

We generated atomistically resolved amorphous thin film morphologies using the Metropolis Monte Carlo based simulation method Deposit.[21,22] This method mimics the vapor deposition process by subsequently adding molecules to a simulation box where they can explore the energy landscape on the substrate formed by their predecessors. The interaction between molecules is described using a model including electrostatic point charge and Lennard-Jones interactions. Intramolecular degrees of freedom apart from dihedral angle rotations are frozen to reduce the effect of dynamic disorder.[23] The torsion potentials as well as the partial charges are parameterized in a molecule specific way using density functional theory (DFT) calculations (B3-LYP/def2-SV(P) level of theory).[24–26]

Bulk energy levels were calculated using the Quantum Patch method that self-consistently equilibrates the electronic structure of large molecular systems using an electrostatic embedding model.[27–29] The Quantum Patch method furthermore allows to charge selected molecules in a system, which enables the calculation of bulk electron affinities and bulk ionization energies using the ΔSCF procedure.[30]

To study the influence of water molecules on the density of states, we generated 200 clusters of single water molecules surrounded by α-NPD using Deposit. We analysed the clusters using Quantum Patch to quantify the shift of α-NPD frontier orbital energies due to the electrostatic interaction with the strong dipole moment of water (2.15 Debye in DFT, B3-LYP/def2-SV(P) level of theory). Kinetic Monte Carlo simulations[31–34] of hole and electron transport in α-NPD systems containing water molecules were performed to analyse the water concentration dependent charge carrier mobility. We used stochastically upscaled boxes of 1,000,000 α-NPD molecules.[35] Gaussian energy disorder parameters and distance dependent electronic couplings were calculated using the Quantum Patch method. The influence of the water molecules was modelled by stochastically adding the precomputed shifts caused by the water molecules at varying water concentrations.

To analyse the electron affinity and trap depth of oxygen molecules and water-oxygen clusters, we generated mixed morphologies of the host materials α-NPD, tris(4-carbazoyl-9-ylphenyl)amine (TCTA)[36] and 1,3,5-tris(N-phenylbenzimidazol-2-yl) benzene (TPBi)[37] and the impurities $O_2$, $O_2$-$H_2O$ and $O_2$-$(H_2O)_2$. In contrast to the simulations described in the previous paragraph, we are here interested in absolute energy levels and traps depths. For this reason, initial geometries of all molecules in vacuum (neutral and ionized states) were optimized at a B3-LYP/dev2-TZVP level of theory for all host molecules and a B3-LYP/def2-QZVP level of theory and a spin-multiplicity of 3 (triplet) for all impurities, followed by B3-LYP/def2-QZVP single point calculations to obtain adiabatic vacuum ionization potentials and electron affinities using the ΔSCF procedure.[38] To benchmark this procedure, we compared our results to CASPT2/aug-cc-pv5z calculations of $O_2$ molecules in vacuum and obtained similar values of electron affinity (0.20 eV in CASPT2 and 0.26 eV in our work, compared to 0.4 eV in experiment [39]) and bond length (1.20 Å / 1.35 Å in CASPT2 vs. 1.20 Å / 1.35 Å in B3-LYP).[40] We then used the Quantum Patch method to calculate static and dynamic polarization effects on the electron affinities of 100 host molecules and 25 impurity molecules in each host-impurity combination. We observe a convergence of the bulk electron affinity and ionization potential of both host and impurity molecules after ~5-6 iterations of the Quantum Patch method and when taking into account ~100 neighboring molecules as polarized embedding size (see **Figure 5c** and **5d**).

### 3.1 Shallow traps and exponential density of states

Water molecules are omnipresent in ambient conditions and are thus a likely candidate to cause trap states in organic thin films. However, we find that the electron affinity of water molecules (see Table 1) is low, which makes it unlikely that the water molecules themselves act as traps. Yet, their large permanent dipole moment of 2.15 Debye can cause electrostatic disorder when interacting with other molecules in thin films. In order to quantify this effect, we generated 200 clusters, each with 30 α-NPD molecules surrounding a single water molecule. Using the Quantum Patch method, we calculated the shift of the HOMO and LUMO energies of each α-NPD molecule in the presence of the water molecule.

**Figure 2** shows these energy level shifts as a function of the nearest-atom distance between the α-NPD and the water molecules. The shifts of the LUMO level are on average larger than the shifts of the HOMO levels. Strongest (negative) LUMO shifts occur when the water molecule is in proximity to the naphthalene group of α-NPD, with the hydrogen atoms pointing towards the naphthalene group. Analysis of the HOMO and LUMO orbitals of randomly selected α-NPD conformers (see **Figure 3c** and **3d**) shows that the LUMO orbital of α-NPD is mostly localized on the naphthalene groups, leading to a strong electrostatic interaction between the LUMO and the dipole moment of the water molecules. The HOMO orbitals are more delocalized along the backbone of the molecule, leading on average to weaker interactions with the dipole moments of water molecules.

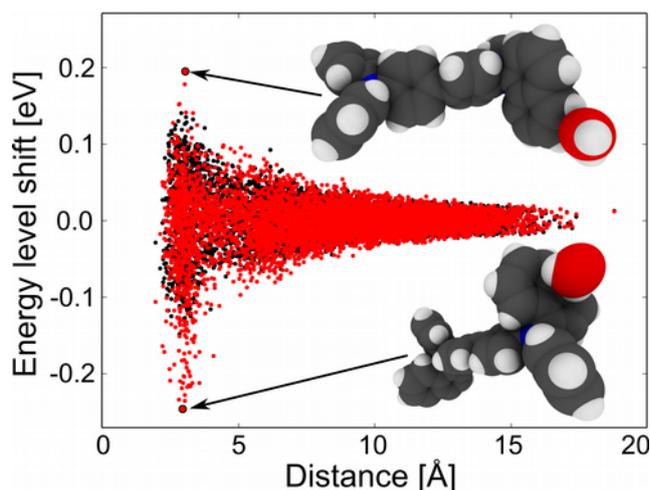

*Figure 2.* HOMO (black) and LUMO (red) energy shifts of α-NPD molecules as a function of the nearest-atom distance to a water molecule. The inset shows the α-NPD/water conformers that caused the strongest (positive and negative) LUMO level shifts.

To quantify the influence of water molecules on the hole and electron mobilities of amorphous α-NPD films, we performed kinetic Monte Carlo simulations on stochastically upscaled periodic disordered systems[35] and calculated the charge carrier mobility as a function of the applied electric field and the concentration of water molecules. For pristine α-NPD films, we assumed a Gaussian density of states with disorder parameters of $\sigma_e$ = 0.075 eV (electrons) and $\sigma_h$ = 0.087 eV (holes).[41–43] The reorganization energies of α-NPD are $\lambda_e$ = 0.109 eV and $\lambda_h$ = 0.158 eV (B3-LYP/def2-TZVP) and the distance dependence as well as the (off-diagonal) disorder of transfer integrals were calculated using the Quantum Patch method.

To model the presence of a given concentration of water molecules in the α-NPD films, we stochastically selected a certain fraction of α-NPD molecules and changed their energy levels with randomly selected energy level shifts caused by water molecules with a nearest-atom distance of <4 Å (see **Figure 2**). This procedure assumes that each water molecule only interacts with the nearest host molecule and thus underestimates the full impact of water molecules on the density of states of the host material by neglecting the long-range coulomb interactions. At the same time, it avoids the introduction of incorrect correlation effects that might be caused by randomly assigning energy level shifts to more than one host molecule per impurity.

The impact of the presence of up to 1 wt% water molecules (corresponding to 25 mol%) on the density of states of the α-NPD host material is illustrated in **Figure 3a**. These artificially high concentrations were chosen to illustrate the impact on the density of states. As shown in **Figure 3c**, the impact on the (electron) mobility is also significant at lower water concentrations. While the density of states for holes (HOMO distribution) is slightly widened (compared to the Gaussian density of states shown as a dashed line), the density of states of electrons shows the formation of exponential tail states, in particular towards lower energies. This asymmetry of holes and electrons is a consequence of the asymmetry of energy level shifts shown in **Figure 2**, caused be the stronger localization of α-NPD LUMO levels compared to α-NPD HOMO levels (see **Figure 3c** and **3d**).

Using this density of states in kinetic Monte Carlo simulations, we observe a decrease in hole mobility of approximately a factor of two, while the electron mobility decreases by more than one order of magnitude. This model assumes a static orientation of the water molecules on the time scale of charge hopping processes. In reality, the orientation of the water molecules randomly fluctuates on a characteristic ps timescale, changing the magnitude and even sign of the energy level shifts shown in **Figure 2**. However, the presence of charge carriers on nearby molecules will influence the orientation of the water molecules due to charge-dipole interactions, favoring an orientation where the water dipole stabilizes the charge and thus creates a shallow trap. As shown in **Figure 2**, only molecules within a center-of-mass distance of 5-10 Å to the closest water molecule experience significant energy level shifts. Thus, while dynamic effects might change the energy landscape experienced by the charge carriers, only molecules close to water impurities can potentially become shallow traps.

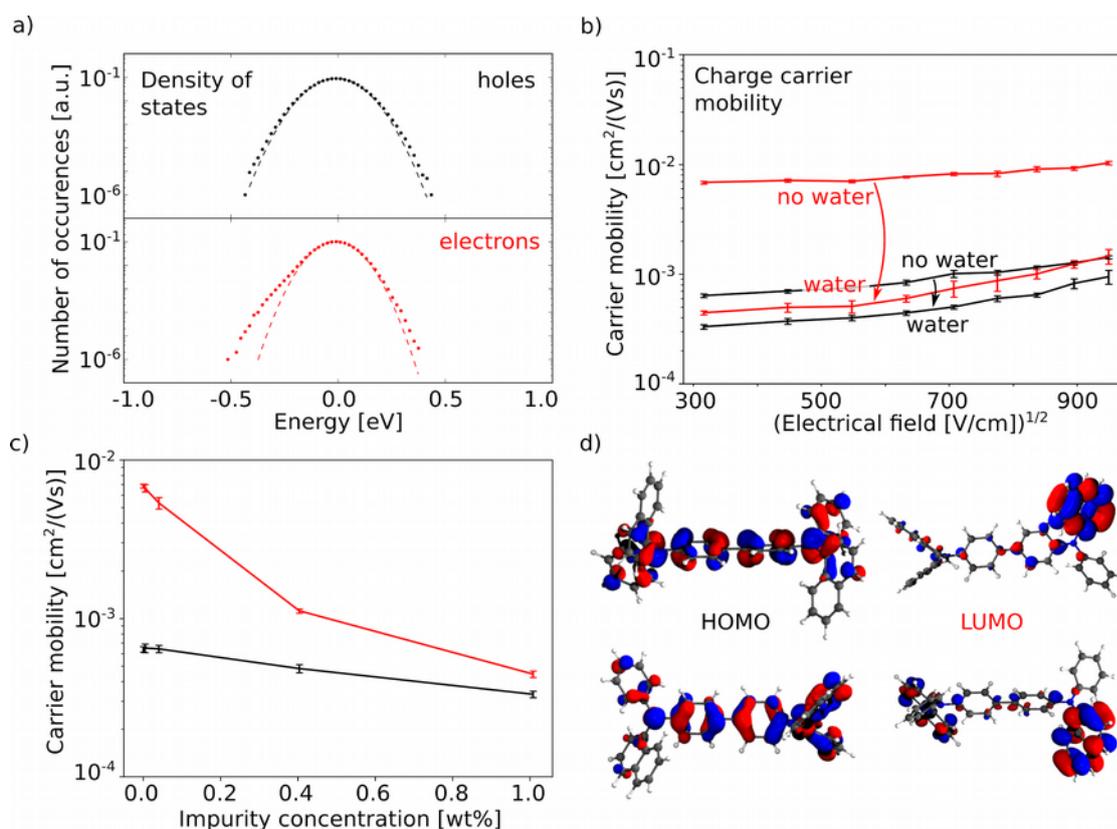

*Figure 3.* a) The presence of water molecules in α-NPD causes the introduction of exponential tails on the density of states. b) kinetic Monte Carlo simulations of the hole and electron mobility as a function of applied electric field. The exponential tail states caused by the presence of water molecules result in a decrease of the charge carrier mobility, in particular the electron mobility. c) and d) α-NPD HOMO and LUMO for different randomly selected conformers.

### 3.2 Deep traps caused by oxygen molecules and oxygen-water complexes

While water molecules and other molecules with intrinsic dipole moments can cause a broadening of the density of states and even result in exponential tail states (shallow traps), they cannot explain the experimental observation of deep traps in the band gap and the universal trap level found in many organic semiconducting materials.[5,9,44] For this reason, we performed a systematic search for potential impurities that can explain the experimental findings. A universal trap level as found in Nicolai *et al.*[5] can hardly be explained by impurities such as side products, non-reacted educts or (parts of the) catalysts. We thus focussed on abundant small molecules that are omnipresent under ambient conditions and that cannot fully be excluded during device fabrication.

To quantify their potential for active trapping of electrons, we calculated their electron affinity in vacuum at a B3-LYP/def2-QZVP level of theory (that turn out to be in good agreement with CASPT2/aug-cc-pv5z reference calculations [40,45] and experimental data of the $O_2$ molecule, see **Methods**). The resulting vacuum electron affinities of ten potential impurity candidates are shown in **Figure 4** and **Table S1**. $O_2$ and $CO_2$ show intrinsic vacuum electron affinities of 0.255 eV (0.4 eV in experiment [46]) and -0.655 eV, respectively. Combining these molecules to $O_2$-water and $CO_2$-water complexes significantly increases their vacuum electron affinities to maximally 2.006 eV for a $O_2$-$(H_2O)_2$ complex.

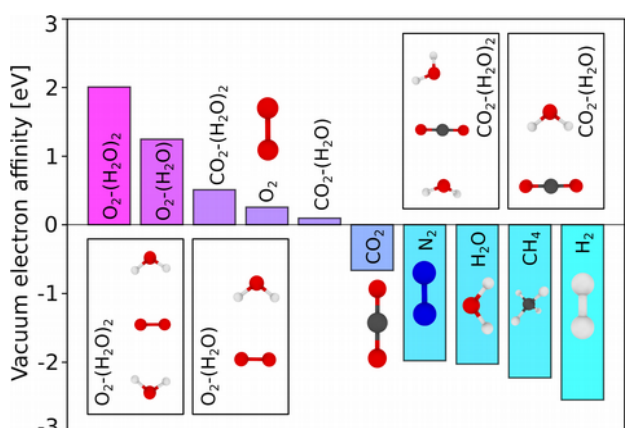

***Figure 4.*** *Vacuum electron affinities of potential impurities. The highest electron affinities are observed for $O_2$, $O_2$-water complexes and $CO_2$-water complexes.*

To compare the electron affinity of these potential electron trapping molecules to commonly used organic semiconductors, bulk effects such as electrostatic polarization and dynamic screening of charges have to be taken into account. For that reason, we created amorphous morphologies of three commonly used host materials with varying LUMO levels mixed with three impurities, namely $O_2$, $O_2$-$(H_2O)$ and $O_2$-$(H_2O)_2$. We used the Quantum Patch method to calculate the bulk electron affinities of randomly chosen host and impurity molecules in these amorphous bulk systems.

**Figure 5a** and **b** show a comparison of the vacuum electron affinities and the bulk electron affinities of all studied material combinations. In the vacuum calculations shown in **Figure 5a**, only $O_2$-water complexes have electron affinities larger than those of the host materials. In bulk simulations (**Figure 5b**) we observe a significant broadening of the distributions of electron affinities of both host and impurities and a shift towards deeper energies. As shown in **Figure S1**, the energy level shift of a host molecule (*e.g.* α-NPD) due to electrostatic polarization of its environment is significantly weaker than that of an impurity molecule (*e.g.* $O_2$). However, potentially due to the stronger electrostatic interaction between the highly localized additional charge on the impurity molecule, the polarization effect is stronger in case of $O_2$ compared to α-NPD, which leads to shifts of more than 2 eV in electron affinity and ionization potential. Previous studies assumed a value of 1.5 eV, independent of the type of impurity.[5]

Taking into account polarization effects we observe a wide distribution of trap levels and trap depths that can explain the experimentally observed universal trap level on the basis of $O_2$ molecules alone, without the need to consider water-oxygen complexes. This observation also explains the large spreading in the magnitude of the electron current in various amorphous organic semiconductors. As shown in **Figure 6** the electron current in TPBi is several orders of magnitude higher as compared to α-NPD and TCTA. This is opposite to what is expected when considering the theoretically calculated electron mobilities of α-NPD and TCTA, which are even higher than their hole mobilities. In experiment, the electron currents in α-NPD and TCTA are both very low, showing that severe charge trapping hinders the transport. If charge trapping was only the result of a broadening of the DOS distribution, a less severe reduction of the electron transport would be expected. However, we observe that the shift in the oxygen impurity electron affinity is independent of the host material. As a result, molecular-oxygen induced traps lead to shallower traps in case of the TPBi host material compared to α-NPD and TCTA. The electron currents can be described using a numerical device model containing electron traps that are Gaussian distributed in energy.[5] We find that the amount of electron traps $N_t$ equals $2.6·10^{23}$ m$^{-3}$ for TPBi, $1.1·10^{24}$ m$^{-3}$ for α-NPD and $7.8·10^{23}$ m$^{-3}$ for TCTA, respectively. As a result, all trap concentrations are in the range between $10^{23}$ m$^{-3}$ and $10^{24}$ m$^{-3}$ as has been reported before for various organic semiconductors.[5] The trap depth $E_t$ on the other hand amounts to 0.4 eV for TPBi, 0.67 eV for α-NPD and 0.8 eV for TCTA, measured between the centers of the LUMO DOS and trap DOS distributions.

Our multiscale simulations thus reveal the origin of the enhanced electron transport in typical electron transport materials, such as TPBi. The presence of molecular oxygen results in a universal electron trap with a reduced trap depth. In contrast, the strongly reduced electron transport in hole transport materials as TCTA and α-NPD is the result of deep electron traps.

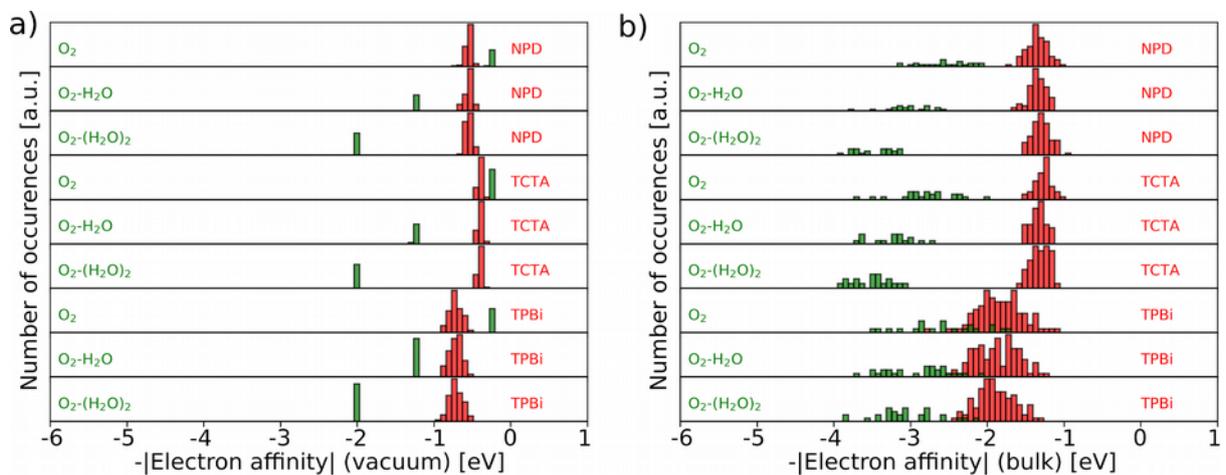

*Figure 5.* a-b) Distribution of (negative) electron affinities of host (α-NPD, TCTA and TPBi) and impurity molecules in a) vacuum and b) bulk calculations. While $O_2$ does not act as an electron trap in vacuum, it becomes a trap in thin films.

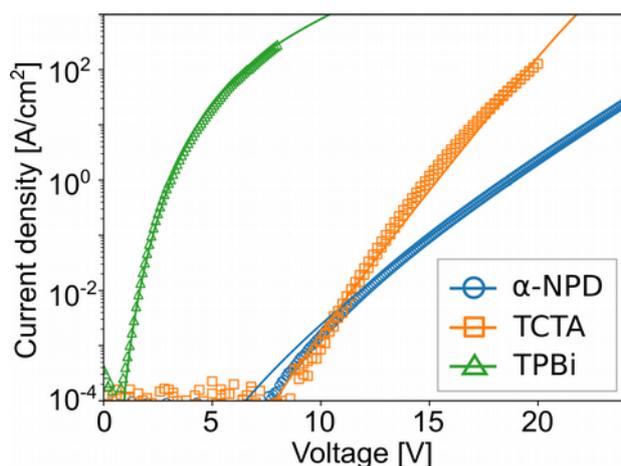

*Figure 6. Current density–voltage characteristics of electron-only devices for TPBi, TCTA and α-NPD, with layer thicknesses of 115 nm, 110 nm, and 111 nm, respectively. The device structure is glass/Al/ organic semiconductor/Ba/Al.* Experimental values are shown as symbols, simulated data is shown as lines.

To analyse the influence of the trap states on electron transport in α-NPD, TCTA and TPBi, we performed KMC calculations of all three systems in presence of a varying concentration of molecular $O_2$ traps. We used mean EA values of 1.35 eV, 1.27 eV and 1.84 eV (see **Figure 5b**), energy disorder parameters of 93 meV, 98 meV and 115 meV, reorganization energies of 145 meV, 75 meV and 343 meV for α-NPD, TCTA and TPBi, respectively. Reorganization energies of all materials as well as energy disorder parameters were computed using B3-LYP/def2-SV(P) calculations in vacuum. Energy disorder parameters of α-NPD and TCTA were computed using pristine Deposit morphologies and the Quantum Patch method and averaged over three independent simulation boxes per material. For TPBi, this method overestimates the energy disorder, probably due to contributions of dynamic disorder. We therefore fitted the energy disorder parameter to obtain a trap-free mobility of $10^{-5}$ cm$^2$/Vs. We used a trap level of 2.6 eV with a standard deviation of 0.4 eV for molecular oxygen in all host materials. This leads to trap depths slightly higher than observed in experiment but consistent with the multiscale simulations shown in **Figure 5**. The field strength was set to $3.6 \cdot 10^5$ V/cm, the relative permittivity to 4.0, the temperature to 300 K and the density of electrons to $10^{-4}$ electrons per nm$^3$. We used an upscaled, periodic, amorphous morphology of size 100 nm X 100 nm X 100 nm with 1.13 molecules/nm$^3$, 0.88 molecules/nm$^3$ and 1.04 molecules/nm$^3$, respectively, derived from the atomistic morphologies generated in this study. Electron-electron interactions were treated explicitly in the KMC protocol.

We find a strong deterioration of the electron mobility of the host materials when the trap concentration is approaching the electron concentration (see **Figure 7**). To quantify this effect, we computed the change in electron mobility from low trap concentration to the point where trap and electron concentration are the same. We observe a reduction of the electron mobility in α-NPD/TCTA by a factor of 28/39, while the electron mobility of TPBi is only reduced by a factor of 6.1. The deep traps in α-NPD and TCTA trap all electrons and significantly reduce the electron mobility to a point where almost no transport is visible in the KMC simulations. In case of more shallow traps in TPBi, not all electrons are trapped and transport is still measurable beyond the point of equal electron and trap concentration.

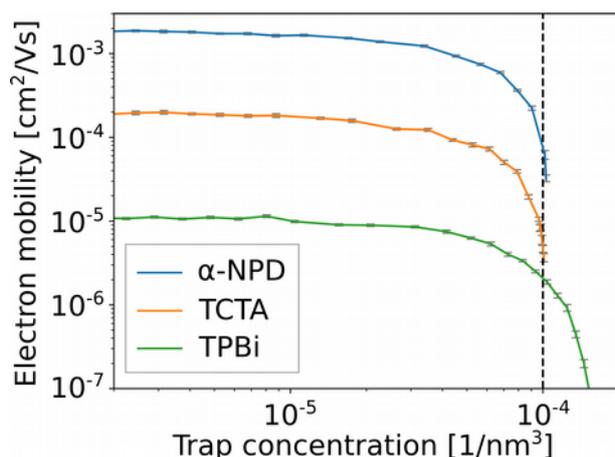

*Figure 7. Electron mobilities of α-NPD, TCTA, TPBi in the presence of a variable concentration of $O_2$ traps. The vertical dashed line indicates the electron concentration. A steep decline of the electron mobility of all materials can be observed when the trap concentration approaches the electron concentration (vertical dashed line). However, in case of deep traps (α-NPD and TCTA), transport is completely blocked by traps whereas in case of shallower traps (TPBi), electrons still move across the organic layer at trap concentrations higher than the electron concentration.*

## 4. Conclusions

We used a multiscale simulation method to investigate the influence of water molecules on the density of states of host materials. We furthermore quantitatively analysed the electron affinity of a series of small inorganic molecules to assess their potential for acting as active electron traps in amorphous organic semiconductors.

In agreement with Borsenberger/Bässler theory,[47] our simulation results suggest that shallow traps and exponential tail states form in the presence of water molecules or any other molecule with a permanent dipole moment. Depending on the concentration of water molecules in thin films, this can have a significant impact on the charge carrier mobility of the host material. In principle, the effect is independent of the host material. However, host materials with localized orbitals such as the LUMO orbital of α-NPD are more affected because of stronger electrostatic interactions between the orbital and the dipole moment of the impurity. Molecules that sustain spatially delocalized frontier orbitals might help to mitigate this effect. However, due to the stronger response of delocalized orbitals to the molecular conformation, such molecules might also tend to show higher values of (conformational) energy disorder, which is one of the reasons why the intrinsic electron mobility of α-NPD is higher than the hole mobility.

According to our simulations of impurities that can potentially act as active traps, molecular oxygen and its complexes with water molecules can cause deep traps and explain the experimentally known universal trap level. The concentration of water-oxygen complexes (O($c_o \cdot c_w$)) is significantly lower than the concentration of molecular oxygen ($c_o$), making the latter the most likely candidate for deep traps. We furthermore find that the trap depths caused by molecular oxygen are closer to the experimentally observed trap depths in α-NPD, TCTA and TPBi. In agreement with experiment, we find deeper traps in the case of α-NPD and TCTA due to their lower electron affinities, while the trap depth in TPBi is reduced due to a higher host electron affinity. Charge transport simulations show that traps in α-NPD and TCTA effectively trap all free electrons, leading to a steep decline in electron mobility when the trap concentration approaches the electron concentration, which is not the case in TPBi.

We hope that the modeling approach applied here will contribute bridging electronic processes going on at the microscopic scale, such as local interactions with impurities, to macroscopic charge transport in amorphous organic semiconductors.

**Acknowledgement**
P.F. acknowledges funding by the DFG grant no. 398357793 and from the European Union's Horizon 2020 research and innovation programme under the Marie Skodowska-Curie grant agreement no. 795206 (MolDesign). D.B. is a FNRS Research Director. This project has received funding GRK 2450 Scale bridging methods in computational nanoscience from the DFG and from the European Union Horizon 2020 research and innovation programme under grant agreements no. 646176 (EXTMOS) and no. 646259 (MOSTOPHOS). This work was performed on the computational resources ForHLR I and ForHLR II funded by the Ministry of Science, Research and the Arts Baden-Württemberg and DFG ("Deutsche Forschungsgemeinschaft").

# Supporting Information

### 1. Electron affinities of potential trap materials

**Table S1** shows the DFT (B3-LYP/def2-QZVP) calculated electron affinities (delta SCF method) of various potential impurtiy candidates.

*Table S1: Comparison with other potential traps: $N_2$, $CO_2$, $H_2$, $CH_4$, $H_2O$, $H_2O$-dimer → no comparable electron affinity as $O_2$.*

| Material | DFT vacuum EA |
|---|---|
| $O_2$ | 0.255 |
| $O_2$-$H_2O$ | 1.245 |
| $O_2$-$(H_2O)_2$ | 2.006 |
| $CO_2$ | -0.665 |
| $CO_2$-$H_2O$ | 0.097 |
| $CO_2$-$(H_2O)_2$ | 0.510 |
| $N_2$ | -1.980 |
| $N_2$-$H_2O$ | -1.103 |
| $N_2$-$(H_2O)_2$ | -0.438 |
| $H_2$ | -2.550 |
| $CH_4$ | -2.229 |
| $H_2O$ | -2.029 |
| $(H_2O)_2$ | -1.119 |

## 2. Vacuum energy levels compared to bulk energy levels

**Figure S1** shows the IP and EA levels of an α-NPD molecule and an $O_2$ molecule as computed using the Quantum Patch method using different embedding sizes. We observe a convergence of the energy levels after few (4-7) iteration steps. Approximately 100 molecules as polarizable embedding size are required to converge the bulk IP and EA levels. The difference between vacuum and bulk IP/EA levels is larger in case of $O_2$ compared to α-NPD.

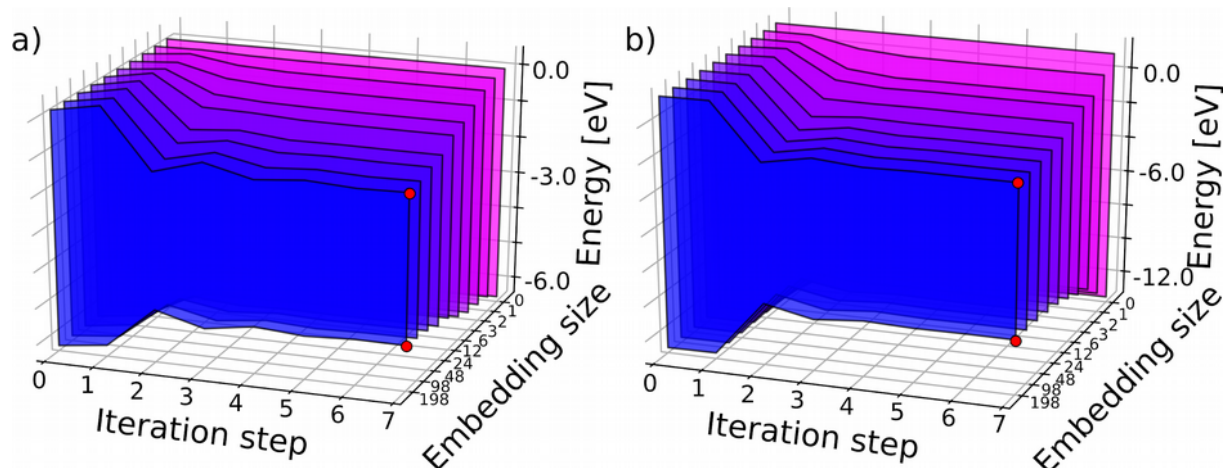

*Figure S1. IP and EA levels of an α-NPD molecule (a) and an $O_2$ molecule (b) as a function of the embedding size (number of molecules) the iteration step of the Quantum Patch method, which represents a gradual polarization of the environment due to the additional charge in the anionic (EA) and cationic (IP) state in the ΔSCF procedure.*

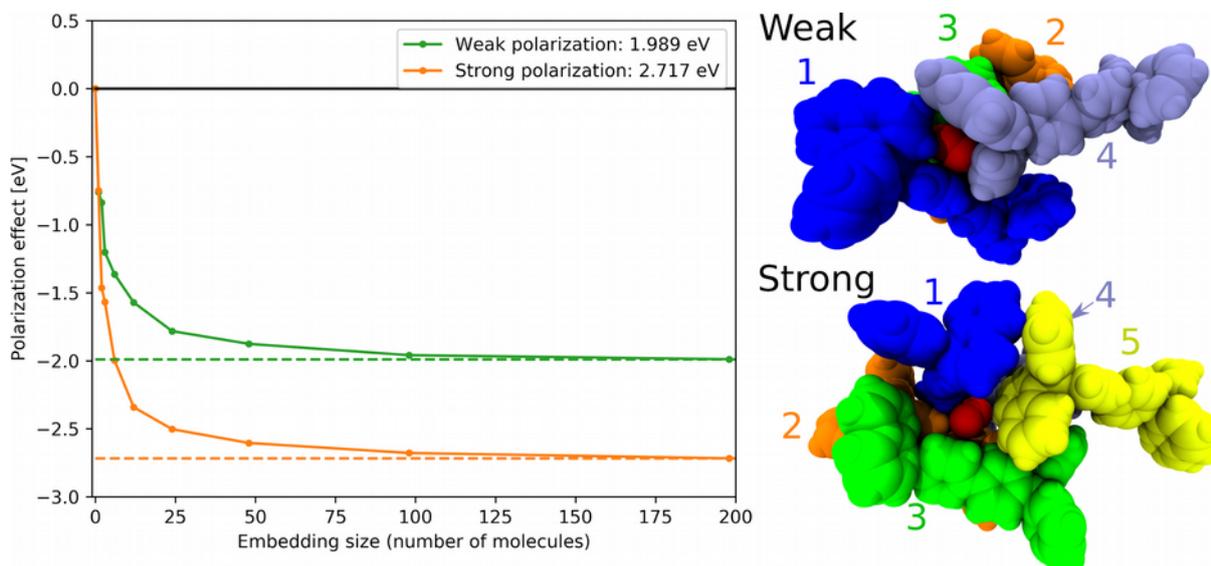

*Figure S2:* Comparison between the polarization energy of two O2 molecules (red), one with a low (1.99 eV) and one with a high polarization energy (2.72 eV). Embedding molecule 1 have similar polarization effects (0.75 eV) in both clusters, whereas molecule 2 has a much stronger polarization effect in the second cluster. After embedding the O2 molecules in 3 α-NPD molecules, the polarization energy increases to 1.20/1.57 eV. The long range polarization effects due to α-NPD molecules 12-198) are very similar (0.21 eV in both cases), indicating that the difference between the two clusters arises from differences in short range interactions. The total polarization effect eventually leads to a bulk electron affinity of -2.25 eV and 2.98 eV after adding the vacuum electron affinity. (see α-NPD/$O_2$ distribution in Figure 5b)